\documentclass[12pt, a4paper]{article}
\usepackage{graphicx}

\begin{document}
\title{An adaptive Metropolis-Hastings scheme: sampling and optimization} 
\author{David H. Wolpert$^{ 1}$\thanks{dhw@email.arc.nasa.gov}
\ and \
Chiu Fan Lee$^{2}$\thanks{c.lee1@physics.ox.ac.uk}
\\
\\
$^1$NASA Ames Research Center\\
MailStop 269-1, Moffett Field, CA 94035-1000 \\ \\
$^2$Physics Department\\ Clarendon Laboratory, Oxford University\\
Parks Road, Oxford OX1 3PU, U.K.
}

\maketitle

\abstract{We propose an adaptive Metropolis-Hastings algorithm in which 
sampled data are used to update the proposal distribution.  We use the
samples found by the algorithm at a particular step to form the
information-theoretically optimal mean-field approximation to the
target distribution, and update the proposal distribution to be that
approximatio.  We employ our algorithm to sample the energy
distribution for several spin-glasses and we demonstrate the
superiority of our algorithm to the conventional MH algorithm in
sampling and in annealing optimization.
} 

\newpage

Monte Carlo methods are powerful tool for evaluating integrals and
simulating stochastic systems (for a recent review, see \cite{Liu01}), and 
they have proved very useful for studying thermodynamic properties of model 
systems with many degrees of freedom. The core of any such method is an 
algorithm for producing independently and identically distributed (IID) 
samples of a provided {\it target probability distribution} $\pi(x \in X)$. 
One of the most popular method used is the Metropolis-Hastings (MH) algorithm 
\cite{MH} and many of its variants \cite{new_MH,west93,sims98,gase03}. 
The Markov transition matrix produced by this algorithm is
parameterized by a {\it proposal distribution} $T(x, x')$. Typically
$T$ is set before the start of the Markov chain in a $\pi$-independent
manner and fixed throughout the running of that chain. The rate at
which the associated Markov chain converges to the desired IID sample
is crucially dependent on the relation between $T$ and $\pi$
however. Since the set \{$x(t)$\} produced by the MH algorithm is
(eventually) an IID sample of $\pi$, one can use \{$x(t)$\} to produce
an empirical estimate of $\pi$ \cite{duha00}. This suggests that we
empirically update $T$ along the Markov chain to be an increasingly
accurate estimate of $\pi$. An adaptive version of the MH algorithm
was introduced in \cite{wole04}. Here we investigate that algorithm
and demonstrate its superiority through spin glass experiments.
A full version of this paper can be found at http://tc.arc.nasa.gov/dhw.

\section{Sampling}
For many systems of interest in physics, $X$ is high-dimensional
(e.g., the number of spins in an Ising system), and for the density
estimation of $\pi$ to work well it must be restricted to producing
estimates from a relatively low-dimensional space, ${\cal{Q}}$.
Intuitively, the idea is to try to find the $q \in {\cal{Q}}$ that is
``closest'' to $\pi$ and use that to update $T$, presuming that this
will produce the most quickly converging Markov chain.  We generically
call such algorithms Adaptive Metropolis Hastings (AMH). To specify an
AMH algorithm one must fix the measure of closeness, the choice of
${\cal{Q}}$, and the precise details of the resultant density
estimation algorithm. One must then specify how the estimates of
$\pi(x)$ are used to update $T(x, x')$.

The most popular way to measure closeness between probability
distributions is with the (asymmetric) Kullback-Leibler (KL) distance 
\cite{coth91}:
\begin{equation}
D(p || p') \ \dot{=} \ -\sum_x p(x) \log \frac{p'(x)}{p(x)}
\end{equation}
Recent work in Probability Collectives
\cite{wolp03b} provides insight into how to do
density estimation to minimize KL distance when $\cal{Q}$ is
low-dimensional. In particular, say $\cal{Q}$ is the set of all
product distributions over $X$, $q(x) =\prod_i q_i(x_i)$, which is equivalent 
to the familiar mean-field approximation \cite{OS01}. Then $D(q
|| \pi)$ is minimized if
\begin{equation}
q_i(x_i) \propto e^{
E(\log(\pi) \mid x_i)} \ ,\ \forall i
\label{eq:brouwer}
\end{equation}
which $E(.|.)$ is the expected value of the first entry conditional on the 
fixed second entry \cite{wolp04a}. Accordingly, $D(q || \pi)$ is just the 
associated free energy of $q$ if $\pi$ is a Boltzmann
distribution. 

Here we instead consider $D(\pi || q)$, which it can be argued is more
appropriate, in light of the information-theoretic justification of KL
distance.  The product distribution minimizing this distance can be
written down directly: it has the same marginals as $\pi$, i.e., the
optimal $q$ obeys $q_i =
\pi_i, \; \forall i$ \cite{wolp04a}. Now a Markov chain produced by a
run of the conventional MH algorithm converges to an IID sample of
$\pi$. So the $i$'th component of the elements of that chain,
\{$x_i(t)$\}, become an IID sample of $\pi_i$. Accordingly, those
sample-components give us an IID sample of the distribution minimizing
$D(\pi || q)$.  So if the number of possible $x_i$ values is not too
large, we can use simple histogramming of the elements of the Markov
chain produced by the MH algorithm to form our estimate of each
marginal $\pi_i$, and therefore of the $q$ minimizing $D(q || \pi)$.
This gives us a rule for updating the proposal distribution, $P(T^t
\mid \{x(t)\})$. 

There are a number of subtleties one should account for in
choosing the precise details of $P(T^t
\mid \{x(t)\})$. In practice there is almost always substantial
discrepancy between $\pi$ and $q$, since ${\cal{Q}}$ is a small subset
of the set of all possible $\pi$. This means that setting $T(x, y) =
q(y)$ typically results in frequent rejections of the sample
points. The usual way around this problem in conventional MH (where
$T$ is fixed before the Markov process starts, and therefore is
typically an even worse fit to $\pi$) is to use a ($t$-independent)
$T(x, y)$ that forces $x$ and $y$ to be close to one
another. Intuitively, doing this means that once the Markov chain
finds an $x$ with high $\pi(x)$, the $y$'s proposed by $T(x, y)$ will
also have reasonable high probability (assuming $\pi$ is not too
jagged). We integrate this approach into our AMH algorithm by setting
$T^t(x, y)$ to be $q^t(y)$ ``masked'' and renormalized to force $y$ to
be close to $x$.

Another important issue is that the earlier a point is on the Markov
chain, the worse it serves as a sample of $\pi$. To account for this,
one should not form $q_i(x_i = s)$ at time $n$ simply as the fraction
of all points $t < n$ for which $x_i(t) = s$. Instead we form those
estimates by geometrically aging the importance of the points before
evaluating the fraction. This means that more recent points have more
of an effect on our estimate of $\pi$. This aging has the additional
advantage that it makes the evolution of the proposal distribution a
relatively low-dimensional Markov process, which intuitively should
help speed convergence.

In \cite{sims98,gase03} related ideas of how to exploit
online-approximations of $\pi$ that are generated from the random walk
were explored.  None of that work explicitly considers
information-theoretic measures of distance (like KL distance) from the
approximation to $\pi$. Nor is there any concern to ``mask'' the
estimate of $\pi$ in that work. The algorithms considered in that work
also make no attempt to account for the fact that the early $x(t)$
should be discounted relative to the later ones. In addition, not
using product distributions, parallelization would not be as
straightforward with these alternatives schemes.

\subsubsection*{Our AMH algorithm}
Our proposed algorithm 
consists of three successive phases: the first of these
is the cooling phase and the third is the data collecting phase. In
both of those phases, the conventional Metropolis-Hastings algorithm
is used, i.e., there is no updating on the proposal distribution.  The second 
phase is where the proposal distribution is
adaptively updated. The details are presented below:

Let $N$ be the number of components of $x$ and $q^t$ the estimate of
$\pi$ at the $t$'th step of the walk. We consider the following
algorithm:
\begin{enumerate}
\item Set $T^t(x, y)$ to $q^t(y)$ masked so that $y$ and $x$ differ in
only one component:
\begin{equation}
T^t(x, y) \; \propto \; \delta \left(\sum_{i=1}^N \delta(x_i-y_i)-N+1
\right)
\prod_{k=1}^N q^t_i(y_i)\ .
\end{equation}
\item
As in conventional MH, sample $[0,1]$ uniformly to produce a $r$ and set
\begin{equation}
x(t+1) = \left\{ \begin{array}{ll}
y, & {\rm if} \ r \leq R^t(x(t),y) \\
x(t), & {\rm otherwise} \end{array} \right.
\end{equation}
where
\begin{equation}
R^t(x,y)=\min \left\{ 1 \ ,\ \frac{ \pi(y) T^t(y,x) }{\pi(x) T^t(x,y) }
\right\}.
\end{equation}
\item
{\it Only in phase 2:}\\
Periodically update $q$. If 
${\rm mod}_N(t+1) = 0$, then update the set $\{ q^t_i \}$ by the non-negative 
multiplier $\alpha < 1$: \\
For all $i,x'_i$, if $x'_i=x_i(t)$
\begin{equation}
q^{t+1}_i(x'_i)= \alpha  (q^t_i(x'_i) -1) +1 
\end{equation}
otherwise
\begin{equation}
q^{t+1}_i(x'_i)=\alpha  q^t_i(x'_i)
\end{equation}
If ${\rm mod}_N(t+1) \neq 0$, then
$q^{t+1}_i(x'_i)=q^t_i(x'_i)$.
To avoid {\it freezing} the proposal distribution, $q_i$ is not allowed to get 
too close to the boundary of the probability simplex (i.e., less than 0.2 
$\times$ the initial uniform distribution). 
\item
$t \leftarrow t+1$. Repeat from step 1.
\end{enumerate}
We note again that in the the first phase of the algorithm, $T$ is uniform and 
step 3 is not implemented, in the second phase, all steps above are 
implemented and in the third phase, step 3 is not implemented.
We also note that our updating method depend only on the current state and
hence is much more efficient than previously proposed methods
\cite{sims98}.

\subsubsection*{Sampling Experiments}

Currently there is no consensus on how to quantify ``how close'' a set
\{$x(t)$\} is to an IID sample of $\pi$.  One approach is to input the
set into a density estimation algorithm \cite{duha00}. One can then
use KL distance from that estimated distribution to $\pi$ as the
desired quantification.  This can be problematic in high-dimensional
spaces though, where the choice of density estimation algorithm would
be crucial. However say we have a contractive mapping $F : x \in X
\rightarrow y \in Y$ where $Y$ is a low-dimensional space that
captures those aspects of $X$ that are of most interest.  We can apply
$F$ to the \{$x(t)$\} to produce its image in $Y$, \{$y(t)$\}.  Next
one can apply something as simple and (relatively) unobjectionable as
histogramming to do the density estimation translating \{$y(t)$\} to an
associated estimate of the generating distribution over $Y$. We can
then use KL distance between that histogram and $F(\pi)$ as the
desired quantification of how good our transition matrix is. This is
the approach we took here.

An important issue with KL distance is that the KL distance from (a density of) the
sample points to $\pi$ diverges if no samples are obtained in region
where $\pi$ is substantially non-zero.  This is not a serious problem
if the total probability, $\epsilon$, of such regions KL divergences is
negligible because the discrepany obtained on any expected value
calculations will be bound by $\epsilon$. Figure 2 illustrates
how AMH and conventional MH compare in their values of $\epsilon$.

Our first experiment concerns the Ising spin-glass model:
\begin{equation}
\label{main_model}
H(x)= \frac{1}{2}\sum_{<i,j>} J_{ij} x_i x_j + \sum_i h_i x_i
\end{equation}
where $<i,j>$ denotes summation over all neighbours. In this function
the $J_{ij}$ and $h_i$ are randomly generated integers in the interval
$[-5\ , \ 5]$ and the $x_i$ can take on values $-1$ and $1$.  Our task
is to sample the associated Boltzmann distribution:
\begin{equation}
\pi(x) \propto \exp (-H(x)/T)
\end{equation}
where $T$ corresponds to the temperature in a thermodynamic setting.
We have chosen spin-glasses for illustration because it is generally
believed that they display salient features of complex disordered systems
\cite{spin_glass}.

We have performed experiments on spin-glasses in a 1D ring formation
(with 50, 75 and 100 spins shown in Figure 1).  In these experiments,
we firstly run, with random initial states, 5 long Markov chains
(800,000$\times N$ steps where $N$ is the number of spins and data are
collected at the last quarter of chain) with the conventional MH
algorithm. We then average the energy distributions obtained to form
our {\it target} distribution. Its closeness to the true distribution
is suggested by how small the associated KL distances to the original
distributions are (the bottom three lines in Figure 1).

We then produced 100 samples of energy distributions 
with the MH and the adaptive MH methods, with chains of 40,000$\times N$ steps 
each. We note that in the adaptive MH method, $q_i(x_i)$ corresponds to the 
the probability of spin $i$ being in state $x_i$. 
Data are again collected in the last quarter of each chain in both case,
and we performed proposal distribution updates as detailed before in the
third quarter of the chains in the adaptive M-H case (with the updating 
parameter $\alpha=0.98$). 

Figures 1 and 2 show the results of these experiments with the error bars 
being the errors on the means. We see that AMH (with $\circ$ markers) 
outperforms conventional MH (with * markers) in sampling, as well as in 
avoiding KL divergence. Similar experiments on a 2D lattice have also been 
performed and the
adaptive M-H shows similar superior performance over conventional MH.

\subsubsection*{Optimization Experiments}

We now investigate the performace of using our algorithm for optimization
rather than sampling, i.e., for minimizing energy.  We
consider the same problem as before, with 100 spins. In the simulation
we randomly generate 20 different sets of $\{J,h\}$ for the
Hamiltonian in eq.~\ref{main_model}. The temperature  goes from
1 to 0.05 in 19 equal steps. We produce 50 samples each for the MH and
AMH versions of the algorithm; the results are presented in
Fig.~3.

\section{Conclusion}
We have proposed a new adaptive
Metropolis-Hastings which, with the product distribution assumption, is easy 
to implement in sampling and we have shown its
superiority over conventional Metropolis-Hastings with computer
experiments. 
Compared with adaptive Metropolis-Hastings proposals
\cite{sims98, gase03}, we have demonstrated the usefulness of our
proposed algorithm with highly non-trivial examples, i.e., spin-glasses,
which highlights the usefulness of our proposed algorithm for sampling
complex distribution.
With annealing in temperature, our method is also shown to be useful in 
hard optimization.

\vspace{.2in}
{\bf Acknowledgements:} We thank Bill Macready for stimulating
discussion. C.F.L. thanks NASA, University College
(Oxford) and NSERC (Canada) for finanical support.

\newpage
\begin{figure}
\caption{KL distances between the random walk points and the target
density .(* = MH, $\circ$ = AMH, + = MH with long chains. The error bars are 
errors on the means.) The inset plot shows the percentages of deviation with 
respect to the free energies found with the long chains in the 100-spin system. 
(Error bars are smaller than the markers.) }
\begin{center}
\includegraphics[scale=.7]{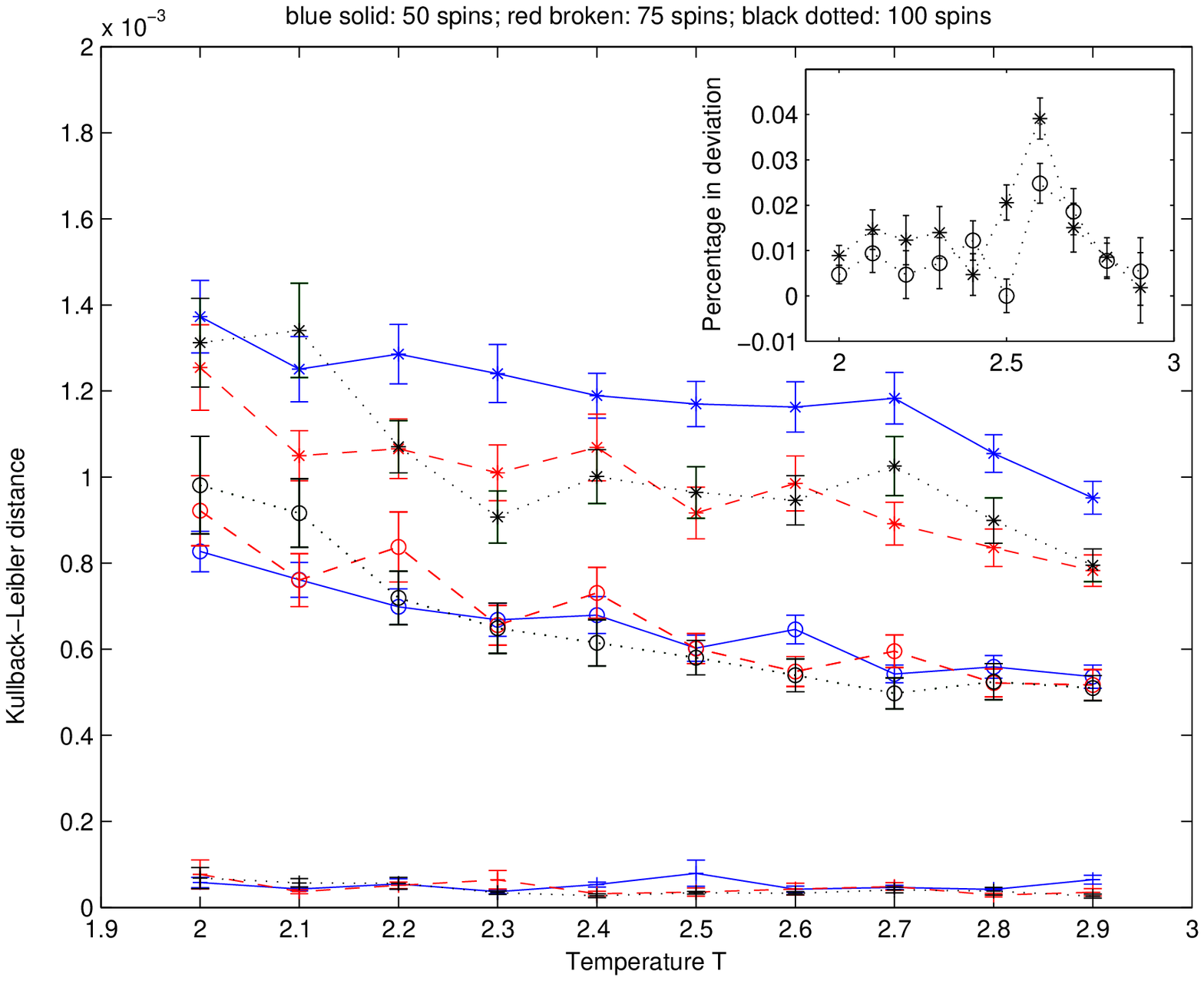}
\end{center}
\end{figure}

\begin{figure}
\caption{(* = MH, $\circ$ = AMH. The error bars are errors on the means.) }
\begin{center}
\includegraphics[scale=.7]{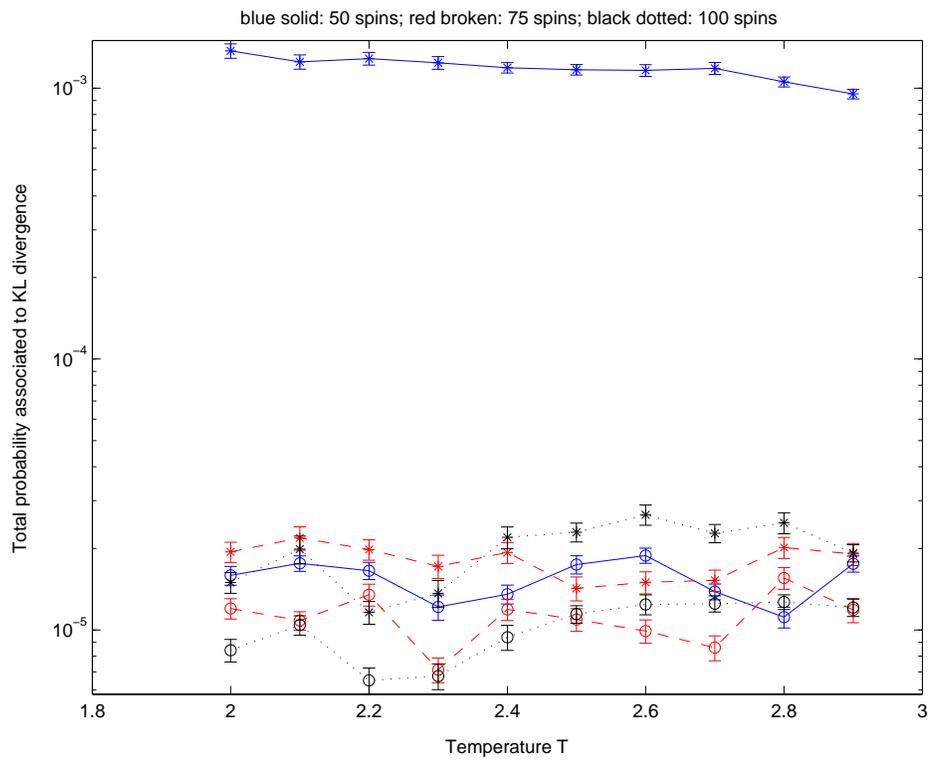}
\end{center}
\end{figure}

\begin{figure}
\caption{Results of 20 different spin-glass experiments. Constants are added to 
the $y$-axis so that the minimum energies found by AMH are 
zero. (* = MH, $\circ$ = AMH. The error bars are errors on the means.) }
\begin{center}
\includegraphics[scale=.7]{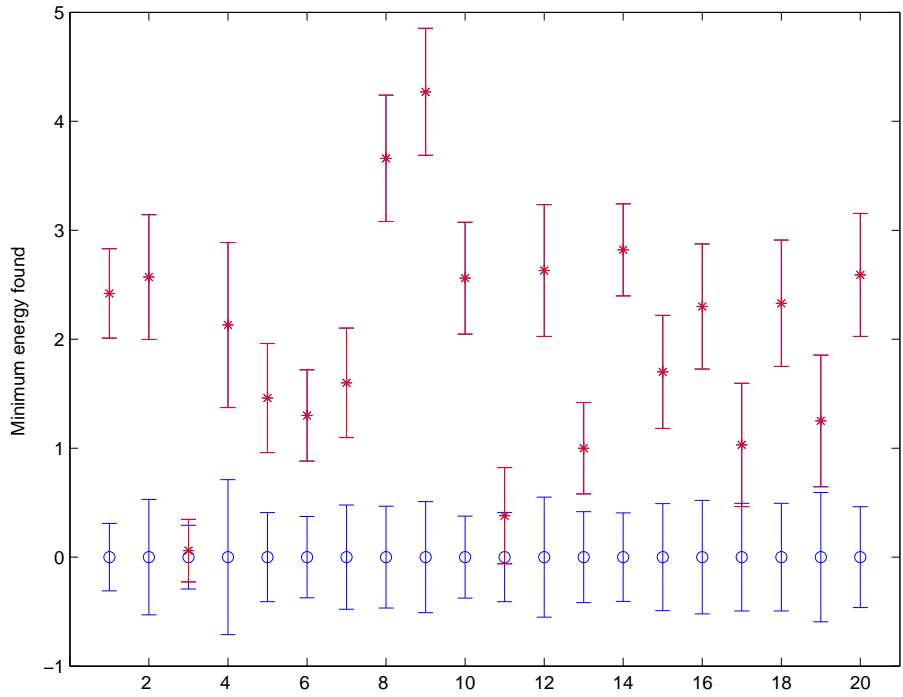}
\end{center}
\end{figure}
\end{document}